\documentstyle[aps,prl,twocolumn,graphicx]{revtex}
\begin{document}

\setlength{\textheight}{24.4cm}

\twocolumn[\hsize\textwidth\columnwidth\hsize\csname    
@twocolumnfalse\endcsname                               

\begin{title} {\large \bf
Spin Correlations in Ho$_{\rm 2}$Ti$_{\rm 2}$O$_{\rm 7}$: 
a Dipolar Spin Ice System}
\end{title}

\author{S.T. Bramwell$^{1}$, M.J. Harris$^{2}$, B.C. den Hertog$^{3}$, M.J.P.
Gingras$^{3,4}$,
J.S. Gardner$^{5}$, D.F. McMorrow$^{6}$,
 A.R. Wildes$^{7}$, A.L. Cornelius$^{8}$, J.D.M. Champion$^{1}$, R.G.
Melko$^{3}$ and T. Fennell$^1$}
\address{$^{1}$Department of Chemistry, University College London, 20 Gordon
Street, London, WC1H OAJ, United Kingdom}
\address{$^{2}$ISIS Facility, Rutherford Appleton Laboratory, Chilton, Didcot,
Oxfordshire OX11 0QX, United Kingdom}
\address{$^{3}$Department of Physics, University of Waterloo, Ontario, Canada
N2L 3G1}
\address{$^{4}$Canadian Institute for Advanced Research, 180 Dundas Street W.,
Toronto, Ontario, Canada M5G 1Z8}
\address{$^{5}$National Research Council, NPMR, Chalk River Laboratories, Chalk River, Ontario, Canada K0J 1J0}
\address{$^{6}$Condensed Matter Physics and Chemistry Department, \mbox{Ris\o} National Laboratory, DK-4000 Roskilde, Denmark}
\address{$^{7}$Institut Laue-Langevin, 156X, 38042 Grenoble Cedex, France}
\address{$^{8}$Department of Physics, University of Nevada, Las Vegas,
Nevada, 89154-4002}
\date{\today}
\maketitle

\begin{abstract}
The pyrochlore material $\rm Ho_{2}Ti_{2}O_{7}$ has been suggested to
show ``spin ice'' behaviour. We present neutron scattering and specific
heat results that establish
unambiguously that Ho$_2$Ti$_2$O$_7$ exhibits spin ice correlations
at low temperature. Diffuse magnetic neutron scattering from
Ho$_2$Ti$_2$O$_7$ is found to be quite well described by a nearest neighbour
spin ice model and very accurately described by a dipolar spin ice model.
The heat capacity is well accounted for by the sum of a dipolar spin ice
contribution and an expected nuclear spin contribution, known to exist in
other Ho$^{3+}$ salts. These results settle the question of the nature of
the low temperature spin correlations in $\rm Ho_{2}Ti_{2}O_{7}$ for which
contradictory claims have been made.
\end{abstract}

\vskip2pc]                                              

\narrowtext

Spin ice materials \cite{Harris1,Ramirez} are
magnetic substances in which the atomic
magnetic moments  obey the same ordering rules  as the hydrogen
atoms in ice, ${\rm H_{2}O}$. They provide a bridge between the simple
statistical mechanics of ice-type models \cite{Lieb},
and the complex behavior of
frustrated magnets \cite{Ramirez2,Gardner,Moessner,Greedan}, with wider
relevance to diverse areas of research such as high
temperature superconductivity and neural networks. Furthermore, relatives of
spin ice such as ${\rm R_{2}Mo_{2}O_{7}}$ (R = rare earth) are a current
focus of attention for their interesting
electronic properties \cite{Taguchi,Katsufuji}. It is therefore important
and desirable to establish the
detailed physics of the simplest spin ice materials, which provide among the
best model systems for the study of frustration and its broad
consequences.

Experiments on ${\rm Ho_{2}Ti_{2}O_{7}}$  provided the original
motivation for  the concept of spin ice \cite{Harris1}.
 In cubic ${\rm Ho_{2}Ti_{2}O_{7}}$,
(space group $Fd \overline{3} m$), the magnetic ${\rm Ho^{3+}}$ ions
occupy a cubic pyrochlore lattice, a
corner-linked array of tetrahedra that is identical to the lattice formed
by the mid-points of the oxygen-oxygen bonds of cubic ice \cite{Anderson}.
The ground state of ${\rm Ho^{3+}}$ is an effective Ising doublet
of local $\langle 111\rangle$ quantization
axis\cite{Harris1,Blote,Mamsurova,Jana,Rosenkranz},
with nearest neighbor ${\rm Ho^{3+}}$ moments experiencing an overall
ferromagnetic coupling that is largely dipolar in origin. Qualitatively,
${\rm Ho_{2}Ti_{2}O_{7}}$ thus approximates the nearest neighbor spin ice
model \cite{Harris1,Harris2,Bramwell}, in
which ferromagnetically-coupled Ising spins on the pyrochlore lattice lie
parallel to the local $\langle 111\rangle$ axes that point towards the centers
of the
tetrahedra. The ground state of this model requires two spins pointing into
and two pointing out of each tetrahedron. Then if spins represent hydrogen
atom displacement vectors, one obtains the ice rules that lead to Pauling's
result for the extensive zero temperature
entropy of the disordered ground state of ice \cite{Anderson,Pauling}.

Neutron scattering experiments on ${\rm Ho_{2}Ti_{2}O_{7}}$ in zero magnetic
field show only broad diffuse scattering down to a temperature \mbox{$T\sim
0.35$ K}, 
consistent with a
disordered spin ice state\cite{Harris1,Harris3}, while low temperature
muon spin relaxation (\mbox{$\mu$SR}) work also finds no evidence for a magnetic
transition\cite{Harris3}. Field-induced ordered states are also consistent with
the spin ice scenario \cite{Harris1,Harris2}. Combined with a ferromagnetic Curie-Weiss
temperature of
$\theta_{CW} = 1.9 \pm 0.1$ K,
${\rm Ho_{2}Ti_{2}O_{7}}$ would thus appear  to be a
prototypical spin ice material. Another candidate for spin ice behaviour is
${\rm Dy_{2}Ti_{2}O_{7}}$ \cite{Ramirez}, for which specific heat measurements
down to $250$ mK
can be integrated to give the residual entropy expected for the
spin ice model. However, none of the measurements
reported in Ref.\cite{Harris1} or Ref.\cite{Ramirez} constitute a proof
that the
spin ice
ground state exists in these materials in zero applied field.
Moreover, a certain
amount of confusion has recently
arisen due to some contradictory observations:
Siddharthan {\it et al.}  suggested that
the specific heat behaviour of ${\rm
Ho_{2}Ti_{2}O_{7}}$ possibly
indicates a transition to a partially ordered state at
$\sim 0.8$ K \cite{Siddharthan},
in disagreement with the neutron scattering and $\mu$SR results
\cite{Harris1,Harris3}. In this Letter we present neutron scattering and
new heat capacity data that unambiguously establish the spin ice nature of
the zero field spin correlations in ${\rm Ho_{2}Ti_{2}O_{7}}$.

The diffuse magnetic neutron scattering from a flux grown
single crystal of ${\rm Ho_{2}Ti_{2}O_{7}}$ was measured in the static
approximation on the PRISMA spectrometer at ISIS. The
crystal was oriented with $[1,-1,0]$ vertical such that the $(h,h,l)$
scattering plane
included the three principal symmetry axes $[1,0,0]$, $[1,1,0]$ and
$[1,1,1]$. Figure 1a shows the scattering pattern at $T\sim 50$
mK. One
of the main
features of the experimental data is the `four-leaf clover'  of intense
scattering
around $(0,0,0)$. There is also strong scattering around $(0,0,3)$ and a
broad region of slightly weaker scattering around  $(3/2,3/2,3/2)$. These
intense regions are connected by narrow necks of intensity
giving the appearance of bow-ties.
The width of the
intense regions indicates short range correlations on the order of one
lattice spacing.
Qualitatively similar scattering has been observed in ice itself \cite{Li}.

To model this data we use the standard expression for the neutron scattered
intensity $I({\bf q})$
\cite{expression}, along
with the Hamiltonian \cite{denHertog,denHertog2,Gingras}:
\begin{eqnarray}
\label{eqn1}
H&=&-J\sum_{\langle ij\rangle}{\bf S}_{i}^{z_{i}}\cdot{\bf S}_{j}^{z_{j}}
\nonumber \\
&+& Dr_{{\rm nn}}^{3}\sum_{i>j}\frac{{\bf S}_{i}^{z_{i}}\cdot{\bf
S}_{j}^{z_{j}}}{|{\bf
r}_{ij}|^{3}} - \frac{3({\bf S}_{i}^{z_{i}}\cdot{\bf r}_{i j}) ({\bf
S}_{j}^{z_{j}}\cdot{\bf r}_{ij})}{|{\bf r}_{ij}|^{5}} \; ,
\end{eqnarray}
where Ising spins ${\bf S}_{i}^{z_{i}}$ of unit length are constrained to
their local $z_i = \langle 1,1,1\rangle$ axes; $J$ is a near neighbor
exchange coupling and $D$ the dipolar coupling. Because of the local
Ising axes the effective nearest neighbor
energy scales are $J_{\rm nn}\equiv J/3$ and $D_{\rm nn}\equiv 5D/3$ \cite{Gingras}.

The near neighbor spin ice model
\cite{Harris1,Bramwell} corresponds to $D = 0$ and $J$  positive
(ferromagnetic). Data were simulated with 3456 spins (${\rm 6\times 6\times
6}$ cubic unit cells) at $T/J = 0.15$ where the model is effectively in an
ice-rules
ground state. The calculated pattern is shown in
Fig. 1b. It successfully reproduces the main features of the experimental
pattern, but there are differences, notably in the extension of the
$(0,0,0)$ intense region along $(h,h,h)$ and the relative intensities of
the regions around $(0,0,3)$ and $(3/2,3/2,3/2)$. Also, the experimental
data
shows much broader regions of scattering along the diagonal directions.
Clearly, the experimental spin correlations do not
reflect a completely disordered  arrangement of
ice-rule states, but some states are favored over
others \cite{denHertog2,Gingras}.

The more complete dipolar spin ice model \cite{denHertog,denHertog2,Gingras}, has
$D_{nn}
 = 2.35$ K, 
fixed by the lattice constant,
and $J_{nn} = -0.52$ K a negative
(antiferromagnetic) parameter determined by fitting the peak temperature
of the electronic magnetic heat capacity (see
below). Spin ice behavior
emerges in this model from the dominant effect of the
long-range nature of
dipolar interactions \cite{denHertog,denHertog2,Gingras},
 and can quantitatively account for the heat capacity
data of
$\rm{Dy_{2}Ti_{2}O_{7}}$ (taken from
Ref.\cite{Ramirez})\cite{denHertog,denHertog2}.
Using single spin flip dynamics, $I({\bf q)}$ was calculated
\cite{expression} 
on a system size of $1024$ spins
(${\rm 4\times 4\times4}$),
at T = $0.6$ K where significant ground
state correlations have developed,
using standard
Ewald summation techniques which
properly handle
infinite summation of dipole-dipole energy terms
(see Refs.
\cite{denHertog,denHertog2}).
The
calculated pattern is shown in Fig 1c.
It captures most details of the
experimental pattern missed by the near neighbor spin ice model in
Fig. 1b,
such as the four intense regions around
$(0,0,0)$, the relative intensities of the regions around $(0,0,3)$ and
$(3/2,3/2,3/2)$ and the spread of the broad features along the diagonal.
Note also the low scattering
intensity at $(2,2,0)$,
consistent with the
experimental low intensity around at
$(2,2,0)$. 
 The approach of $I({\bf q})$ to zero at $q = 0$ is consistent
with a recent observation that the ac-susceptibility of ${\rm
Ho_{2}Ti_{2}O_{7}}$ approaches zero below $0.7$ K \cite{Matsuhira}.

\begin{figure}
\label{fig1}
\rotatebox{0}{\includegraphics[height=20.5cm]{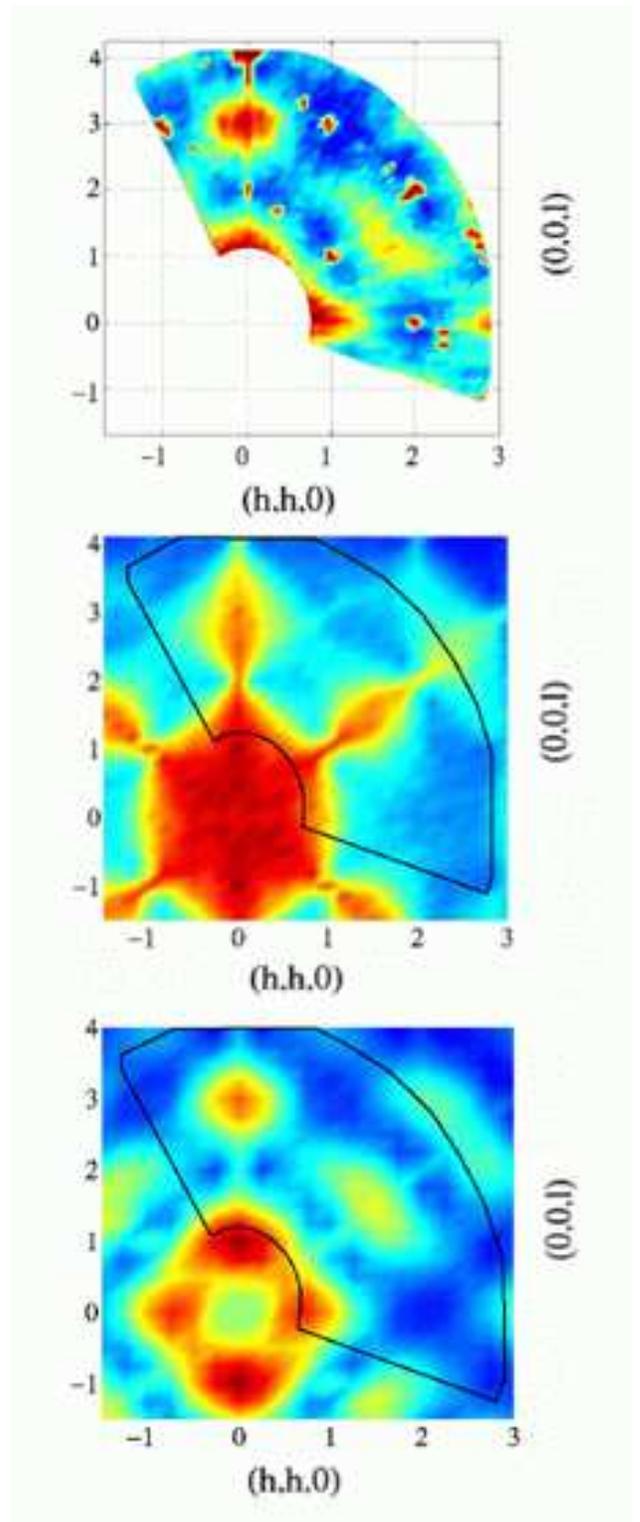}}
\caption{ (a) Experimental neutron scattering pattern of ${\rm
Ho_{2}Ti_{2}O_{7}}$
in the $(hhl)$ plane of reciprocal space at $T\sim 50$ mK.
Dark blue
shows the lowest intensity level, red-brown the highest. Temperature
dependent measurements have shown that the sharp diffraction spots in the
experimental pattern are nuclear Bragg peaks with no magnetic component.
(b) $I({\bf q})$ for the
nearest neighbor
spin ice model at $T=0.15J$.
(c) $I({\bf q})$
 for the dipolar spin ice model at $T = 0.6$ K. The areas defined by the solid
lines denote the experimental data region of (a).}
\end{figure}


The difference between Figs. 1b and 1c shows that dipolar
interactions, whilst inducing a low energy ice-rules manifold, do cause
further
correlations to emerge among  the spins than those that are due to a
nearest neighbor ferromagnetic interaction.
This thermal bias towards
certain ice-rule configurations in the dipolar model is consistent with mean
field calculations\cite{Gingras}, and also with the behavior observed when
single spin flip
thermal barriers are removed by using multiple-spin flip algorithms in
numerical simulations\cite{denHertog2}. In that case, the intense
regions of $I({\bf q})$ evolve into
Bragg peaks upon cooling below $0.17$ K as the system enters a long range
ordered state.
It may
be that at low temperature in ${\rm Ho_{2}Ti_{2}O_{7}}$,
short range correlations
associated with a putative ordered state begin to emerge, but are
dynamically inhibited from developing into Bragg peaks upon further cooling.

To further test the dipolar model, we show in Fig. 2 a comparison between the
calculated $I({\bf q})$
and that measured in a separate experiment with the
same crystal on the IN14 spectrometer at the ILL, Grenoble. Here it was
found that the diffuse scattering was nearly
temperature-independent below 0.7 K. For the theoretical
fit, the form
factor was determined empirically by fitting the neutron data at 35 K
 in the
paramagnetic regime \cite{empirical}. The comparison between theory and
experiment obtained at $0.7$ K along $(0,0,l)$ is shown in Fig. 2a, where it
is seen to be very satisfactory. Hence it is clear that the dipolar
Hamiltonian (1) describes the magnetic behaviour of ${\rm
Ho_{2}Ti_{2}O_{7}}$ to a very good approximation.

\begin{figure}
\label{fig2}
\rotatebox{90}{\includegraphics[height=8cm]{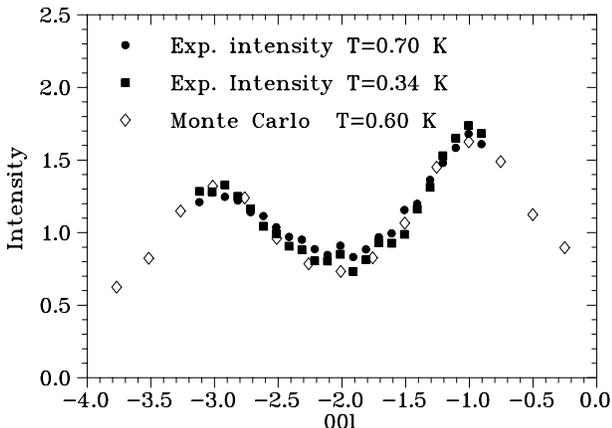}}
\caption{Experimental neutron scattering intensity, $I({\bf q})$,
  of ${\rm
Ho_{2}Ti_{2}O_{7}}$ (filled symbols) along the (00l) direction of
reciprocal space. For quantitative comparison is the intensity (open
 symbols) obtained from Monte Carlo simulation of the dipolar spin ice model
with \mbox{$J_{\rm nn}=-0.52$ K} and \mbox{$D_{\rm nn}=2.35$ K.}}
\end{figure}

\begin{figure}
\rotatebox{90}{\includegraphics[height=8cm]{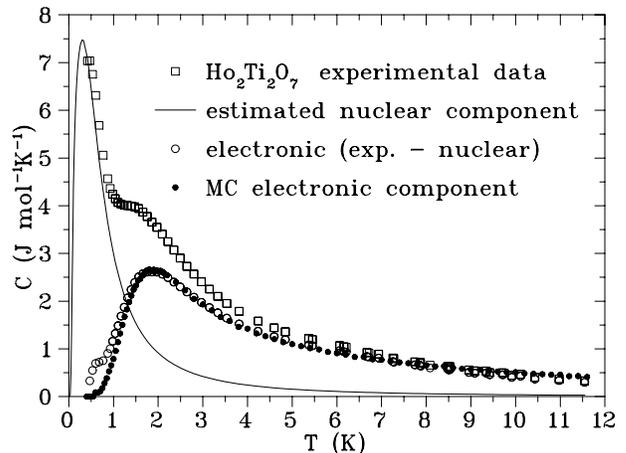}}
\caption{The total specific heat of ${\rm Ho_{2}Ti_{2}O_{7}}$ is shown by the
empty squares. Expected nuclear contribution is indicated by the line,
while the electronic contribution has been estimated by subtracting
these two curves (open circles). Near $0.7$ K this subtraction is prone to
a large error (see text).  Dipolar spin ice simulation results are
indicated by the filled circles.}
\label{fig3}
\end{figure}

We now turn to the heat capacity, analysis of which
is complicated by the previously reported
difficulty in equilibrating this material at low temperatures
($T<0.8$ K)\cite{Siddharthan}.
Single crystals of ${\rm Ho_{2}Ti_{2}O_{7}}$ were synthesized using the floating-zone method 
described in \cite{crystal}. Heat capacity
measurements were performed in a Quantum Design PPMS system inside a 9T
superconducting magnet.  The heat capacity was measured using a thermal
relaxation method between 0.34 and 20~K.  Good thermal contact between the
single crystal and sample stage was ensured with the uses of low temperature
grease.                                                     
Our heat capacity data, shown in Fig. \ref{fig3}, go to lower temperature
than those of Ref. \cite{Siddharthan} and show
no evidence for a phase transition. Rather, there is a noticeable change in
dynamics as equilibration times become longer deep within the
spin ice regime, as signaled by the vanishing electronic specific heat
capacity, but the total heat capacity
continues to steadily increase as the temperature is lowered (note,
however, the shoulder at $T \sim 1.5 K$).

Bl\"{o}te {\it et al.} measured  the heat capacity of the pyrochlore
${\rm Ho_{2}GaSbO_{7}}$ \cite{Blote} and obtained similar behavior to that
found for
${\rm Ho_{2}Ti_{2}O_{7}}$. They successfully
accounted for the heat capacity  below \mbox{$\sim 2$ K} by introducing  a
Schottky anomaly with a theoretical maximum for Ho ($I=7/2$) of $0.9$ R,
due to a splitting of the eight nuclear levels with a level spacing of
$0.3$ K.

The electronic component of the magnetic heat capacity  for
${\rm Ho_{2}Ti_{2}O_{7}}$ may be extracted
by subtracting off the  nuclear
Schottky contribution estimated by Bl\"{o}te {\it et al.}
for
${\rm Ho_{2}GaSbO_{7}}$ \cite{Blote}. This reveals an electronic magnetic heat
capacity that shows  the same characteristic spin ice shape as observed for
${\rm Dy_{2}Ti_{2}O_{7}}$ \cite{Ramirez}. Fig. \ref{fig3} shows a very broad
electronic magnetic  peak centered at \mbox{$T\sim 1.9$ K}.
With this heat capacity revealed,
a natural explanation of the observed change
in dynamics in this material at \mbox{$T\sim 0.8$ K}, (attributed to
a transition in Refs. \cite{Siddharthan,Siddharthan2}),
is that this is
approximately the temperature below which macroscopic relaxation of the
magnetic  degrees of freedom  into the low energy ice-rule manifold has
essentially
ceased (this is also consistent with Ref.(\cite{Matsuhira})).
In other words, the onset of spin ice physics is much higher
than  this temperature, most likely  \mbox{$T\sim 1.9 $ K}.
The behavior of ${\rm Ho_{2}Ti_{2}O_{7}}$ may be compared
to  that of ${\rm Dy_{2}Ti_{2}O_{7}}$, where the onset of spin ice
correlations (peak in the specific heat) is \mbox{$\sim 1.2$ K} and the
magnetic
heat capacity is almost zero at \mbox{$T\sim 0.25$ K}\cite{Ramirez}.

The subtraction of the nuclear contribution, itself an approximation,
becomes prone to error in the range 0.6-0.9 K where the nuclear specific
heat rises sharply. Hence, although our subtraction reveals a small feature
in the electronic specific heat at \mbox{$T\sim 0.7$ K},
a subtraction of the nuclear contribution
from the total heat capacity data reported in Ref. \cite{Siddharthan} 
gives no such feature; thus we conclude that
an electronic-only heat capacity in this region cannot be reliably
determined.

Based on the electronic specific heat peak height, and the temperature at
which it occurs, the dipolar spin ice model allows two independent methods for
determining the value of the nearest neighbor exchange for
${\rm Ho_{2}Ti_{2}O_{7}}$\cite{denHertog}. Using these procedures we find a
nearest neighbor exchange
\mbox{$J_{\rm nn}\sim -0.52$ K}. In Fig. \ref{fig3} we show specific heat
results
from a standard Monte
Carlo single spin flip simulation of the dipolar spin ice model, for a
system size of $6\times 6\times 6$ unit cells
($3456$ spins) with interaction parameters \mbox{$J_{\rm nn}= -0.52$
 K} and nearest neighbor dipole strength \mbox{$D_{\rm nn}=2.35$ K}.
We find quantitative agreement between the experimental electronic specific
heat, $C$,  and that found from the simulation.
At high temperature, $C$
approaches the dipolar spin ice heat capacity
smoothly, as it should since the nuclear contribution is negligible.
At low $T$ it follows the nuclear contribution once the electronic component
has died. Finally,
in-between the low $T$ and high $T$ regimes,
there is the shoulder at $T\sim 1.5$ K in the total
$C(T)$ (alluded to above) corresponding to
a shift of regime between the dipolar spin ice
controlled regime
and the nuclear hyperfine one.

Taken with
our neutron scattering results, it appears clear that
Ho$_2$Ti$_2$O$_7$ is quantitatively
well characterized by the dipolar spin ice model with
a missing entropy close to Pauling's prediction,
and what is found for Dy$_2$Ti$_2$O$_7$ \cite{Ramirez,denHertog}. 
The spin ice behavior revealed explicitly here for ${\rm Ho_{2}Ti_{2}O_{7}}$
is in contrast to the conclusions of Refs. \cite{Siddharthan,Siddharthan2}.
These  were based on a finite size truncation of the dipolar interaction in
the simulations and only a qualitative comparison with experimental
specific heat that neglected the important nuclear contribution (Fig. 3 of
\cite{Siddharthan}).
 Our deduced value of the
antiferromagnetic exchange interaction indicates that
${\rm Ho_{2}Ti_{2}O_{7}}$ is further into the
dipolar spin ice region of the magnetic phase diagram \cite{denHertog}
than is
${\rm Dy_{2}Ti_{2}O_{7}}$, contrary to the claims of Ref.\cite{Siddharthan2},
with
$J^{\rm Ho}_{\rm nn}/D^{\rm Ho}_{\rm nn}\sim -0.22$ and
$J^{\rm Dy}_{\rm nn}/D^{\rm Dy}_{\rm nn}\sim -0.53$.

In conclusion, we have unambiguously established that ${\rm Ho_{2}Ti_{2}O_{7}}$
possesses a spin ice state in zero field. Experiments investigating the
spin correlations of ${\rm
Dy_{2}Ti_{2}O_{7}}$ are in progress \cite{Olegetal}. The Ho salt, in contrast
to its Dy analogue, contains a single rare earth isotope that lends itself
well to neutron scattering and the study of
hyperfine effects. With intriguing spin dynamics and
field-induced ordering phenomena, ${\rm Ho_{2}Ti_{2}O_{7}}$ offers much new
physics to be explored in the field of frustrated magnetism.

M.G. acknowledges financial support from
NSERC of Canada, Research Corporation and the Province of Ontario. 
 We thank B. F{\aa}k, P.C.W.  Holdsworth, and O.A.  Petrenko
for useful discussions.


\begin{references}
\vspace*{-1.5cm}

\bibitem{Harris1}
M. J. Harris, {\it et al.}, \prl  {\bf 79}, 2554 (1997).


\bibitem{Ramirez}
A. P. Ramirez {\it et al.}, Nature {\bf 399}, 333 (1999).

\bibitem{Lieb} E.~H.~Lieb and F.~Y.~Wu,
\newblock{\it Phase Transitions and Critical Phenomena}, ed. C.~Domb and
M.~S.~Green, Academic Press, pp. 332-490 (1972).
G. T. Barkema and M. E. J. Newman, Phys. Rev. E {\bf 57}, 1155 (1998).


\bibitem{Ramirez2}
A. P. Ramirez, Ann. Rev. Mat. Sci. {\bf 24}, 453 (1994).


\bibitem{Gardner}
J. S. Gardner {\it et al.}, \prl  {\bf 82}, 1012 (1999).

\bibitem{Moessner}
R. Moessner, cond-mat/0010301.

\bibitem{Greedan}J.E. Greedan,
``Geometrically frustrated magnetic materials''. To appear in
J. of Mater.
Chem., (2001).


\bibitem{Taguchi} R. Taguchi and Y. Tokura, \prb {\bf 60}, 10280 (1999)

\bibitem{Katsufuji} T. Katsufuji H. \^Y. Hwang and S-W. Cheong, \prl {\bf
84}, 1998 (2000).


\bibitem{Anderson}
P. W. Anderson, Phys. Rev. {\bf 102}, 1008 (1956).



\bibitem{Blote} H.W.J.~Bl\"{o}te {\it et al.},
Physica {\bf 43}, 549 (1969).

\bibitem{Mamsurova} L. G. Mamsurova, K. K. Pukhov, N. G. Trusevich and L.
G. Shcherbakova, Sov. Phys. Solid State, {\bf 27}, 1214 (1985).

\bibitem{Jana} Y. M. Jana and D. Ghosh, \prb {\bf 61}, 9657 (2000).

\bibitem{Rosenkranz}
S. Rosenkranz {\it et al.}, J. Appl. Phys. {\bf 87}, 5914 (2000).


\bibitem{Harris2}
M. J. Harris {\it et al.}, \prl {\bf 81}, 4496 (1998).

\bibitem{Bramwell}S. T. Bramwell and M. J. Harris, J. Phys.: Condens.
Matter {\bf 10},
L215 1998.

\bibitem{Pauling}
L. Pauling, J. Am. Chem. Soc. {\bf 57}, 2680 (1935).

\bibitem{Harris3}
M. J. Harris {\it et al.}, J. Magn. Mag. Mat. {\bf 177-181}, 757 (1998).

\bibitem{Siddharthan}
R. Siddharthan {\it et al.}, \prl {\bf 83}, 1854 (1999).

\bibitem{Li}
J. C. Li {\it et al.}, Phil. Mag. {\bf  B 69}, 1173 (1994).



\bibitem{expression} We use the following expression for the neutron
scattered intensity, $I({\bf q})$:
\begin{eqnarray*}
I({\bf q}) \propto
\vert f(q)\vert ^2 \frac{1}{N}\sum_{ij} e^{{i\bf q.(r_{i}-r_{j})}}\langle {\bf
S}_{i,\perp}^{z_i} \cdot
{\bf S}_{j,\perp}^{z_j}\rangle
\end{eqnarray*}
where $z_i = (1,1,1)$, $\langle\dots\rangle$ denotes a
thermal average and ${\bf S}^{z_i}_{i,\perp}$ is the spin component at
site $i$ perpendicular to ${\bf q}$. The form factor $f(q)$ for Ho$^{3+}$
was calculated in the small-${\bf q}$ approximation;
strictly, an anisotropic form factor for the $M_J = \pm 8$ states of Ho$^{3+}$
should be used, but the error is negligible for the comparison in Fig. 1.

\bibitem{denHertog}
B. C. den Hertog and M. J. P. Gingras, \prl {\bf 84}, 3430 (2000).

\bibitem{denHertog2}
R.G. Melko, B.C. den Hertog, and M.J.P. Gingras
cond-mat/0009225.

\bibitem{Gingras}
M.J.P. Gingras and B.C. den Hertog, cond-mat/0012275.

\bibitem{Matsuhira} K Matsuhira, Y. Hinatsu, K. Tenya and T. Sakakibara, J.
Phys. Condens. Mat., {\bf 12} L649 (2000).

\bibitem{empirical} This procedure also corrects for a small systematic
temperature-independent error in the scattered intensity, believed to arise
from
partial masking of the observed scattering volume. The empirical form
factor varied by 20\%
from the expected one
at $(0,0,3)$.


\bibitem{crystal}
J.S. Gardner, B. D. Gaulin and D. M$^c$K. Paul, J. Crystal Growth,
{\bf 191}, 740 (1998).   

\bibitem{Siddharthan2} R. Siddharthan, B. S. Shastry and A. P. Ramirez,
cond-mat/0009265.

\bibitem{Olegetal} O.A. Petrenko, {\it et al.}, unpublished.

\end{references}
\end{document}